\documentclass[aps,twocolumn,groupedaddress,showpacs, pra, 10pt]{revtex4-2}
\usepackage{graphicx}
\usepackage{physics}
\usepackage{tikz,lipsum,lmodern}
\usepackage[export]{adjustbox}
\usepackage{amsmath}
\usepackage{amssymb}
\usepackage{svg}
\usepackage{graphicx}
\usepackage{natbib}
\usepackage{float}

\usepackage[colorlinks=true, citecolor=blue, urlcolor=blue, linkcolor = magenta]{hyperref}
\usepackage{xcolor}
\usepackage{graphicx}
\usepackage{setspace}
\begin{document}

\title{Quantum Thermodynamics of Open Quantum Systems: Nature of Thermal Fluctuations}

\author{Neha Pathania
}
\email{neha@ctp-jamia.res.in}
\author{Devvrat Tiwari}
\email{devvrat.1@iitj.ac.in}

\author{Subhashish Banerjee}
\email{subhashish@iitj.ac.in}

\affiliation{Department of Physics, Indian Institute of Technology Jodhpur, India}

\begin{abstract}
We investigate the thermodynamic behavior of open quantum systems through the Hamiltonian of Mean Force, focusing on two models: a two-qubit system interacting with a thermal bath and a Jaynes-Cummings Model without the rotating wave approximation. By analyzing both weak and strong coupling regimes, we uncover the impact of environmental interactions on quantum thermodynamic quantities, including specific heat capacity, internal energy, and entropy. Further, the ergotropy and entropy production are computed. We also explore the energy-temperature uncertainty relation, which sets an upper bound on the signal-to-noise ratio. 
\end{abstract}

\maketitle 
\section{Introduction}
The study of thermodynamics within the realm of open quantum systems is a burgeoning area that seeks to understand how classical thermodynamic principles apply when quantum effects are significant ~\cite{miller2018hamiltonian, jahnke2011operational,alicki1979quantum, kosloff}. This field is particularly relevant for systems interacting with their surroundings, leading to complex behaviors and phenomena not observed in isolated systems. 
A fundamental development in the theory of open quantum systems is the Gorini-Kossakowski-Lindblad-Sudarshan (GKLS) master equation~\cite{gorini1976completely,kossakowski1972quantum,breuer2002theory,lindblad1976generators,rivas2012open,banerjee2018open}, which offers a framework for describing the time evolution of a system \(S\) under the influence of weak interactions with a large external environment. Advances in theoretical and experimental domains implore one to go beyond the scope of GKLS dynamics, specifically where strong-coupling non-Markovian effects are prominent~\cite{RevModPhys.89.015001}.

Open quantum systems that are non-Markovian demonstrate the intricate connection between a quantum system and its surroundings. Non-Markovian systems retain a memory of their past, resulting in complex dynamical behavior~\cite{PhysRevLett.103.210401, PhysRevLett.105.050403, PhysRevA.84.052118, RevModPhys.89.015001,doi:10.1142/S1230161218500142,breuer2016colloquium,zhang2012general,liu2011experimental, Utagi2020_SSS, SB_Naikoo_facets, Kumar2018}. A prominent feature of non-Markovian evolution is the resurgence of quantum properties ~\cite{PhysRevLett.103.210401}, the examination of which is crucial for comprehending the system's dynamics under varying levels of coupling with the environment. Studying the dynamics of thermodynamic quantities when the system has strong coupling with the environment in the non-Markovian regime presents a formidable challenge ~\cite{strasberg2016nonequilibrium,thomas2018thermodynamics, PhysRevA.106.032607, SB_Devvrat_Bose_Strong}. 
Recent advances in the study of open quantum systems have yielded significant insights into the behavior of quantum thermodynamic quantities \cite{Lahiri2021, Banerjee2023}, and with technological advancements, this is being leveraged towards quantum thermal devices~\cite{vinjanampathy2016quantum, Kumar2023, Tiwari_Bhanja_2024, PhysRevResearch.2.043302}.
Numerous techniques have been developed recently to address the dynamics of such systems. These encompass the Hamiltonian of Mean Force (HMF)~\cite{PhysRevLett.102.210401,hanggi2008finite,campisi2011colloquium,talkner2020colloquium,hilt2011hamiltonian,strasberg2020measurability}, the method of hierarchical equations of motion, a numerical method utilizing the influence-functional formalism~\cite{tanimura2020numerically,batge2021nonequilibrium}, the reaction coordinate method, which explores strong coupling effects through mapping the Hamiltonian with a reaction coordinate ~\cite{nazir2018reaction, PhysRevA.90.032114,anto2021capturing} and pseudomodes technique~\cite{pleasance2020generalized}. Recently, the reaction-coordinate technique was used in conjunction with the polaron transform to develop an effective Hamiltonian theory dealing with strong-coupling open quantum systems~\cite{PRXQuantum.4.020307}.

Here, we make use of the Hamiltonian of Mean Force to investigate the strong-coupling non-Markovian effects.
This concept of HMF has been previously studied in the Caldeira-Leggett Model~\cite{caldeira1983path}, Spin-Boson Model~\cite{weiss2012quantum, cerisola2024quantum}, Quantum Brownian Motion~\cite{hanggi2005fundamental} to ascertain the system's effective dynamics and thermodynamics as well as in molecular systems~\cite{kapral1999mixed}, and stochastic and macroscopic thermodynamics of strongly-coupled systems~\cite{jarzynski2017stochastic} to study solvation dynamics and chemical reaction rates in complex environments. It has been observed that HMF provides consistent generalizations of the laws of thermodynamics and fluctuation relations~\cite{miller2018energy, PhysRevLett.102.210401}.

The specific heat capacity of the systems coupled to thermal reservoirs demonstrates non-trivial temperature dependencies influenced by quantum effects~\cite{hanggi2008finite, PhysRevLett.102.210401}. Along with ubiquitous entropy, another quantity prominent in quantum thermodynamics is entropy production, which, along with system dynamics, also depends on global state evolution~\cite{landi, esposito2010entropy,deffner2011nonequilibrium}. The Fluctuation Dissipation Relation (FDR), which connects response functions to correlation functions, is fundamental in understanding dissipation and fluctuations ~\cite{kubo1966fluctuation}. 
It turns out that in the strong coupling regime of thermodynamics, FDR has modifications coming from the specific heat capacity and another term that is an essential ingredient in computing the system's internal energy~\cite{miller2018energy, miller2018hamiltonian}. Additionally, the concept of ergotropy, representing the maximum extractable work from a quantum system via unitary transformations, was introduced ~\cite{PhysRevLett.85.1799,allahverdyan2004maximal,alicki2004thermodynamics} and is a crucial ingredient in characterizing the thermodynamic behavior of a system. These studies collectively advance our understanding of the complex behavior and properties inherent to open quantum systems.

The concept of thermal fluctuations has become fundamental in quantum thermometry owing to advancements in technology at the nanoscale that allow temperature sensing at sub-micron scales~\cite{pekola_2006_thermometry, Zanardi_2008, DePasquale2016, carlos2016thermometry, PUGLISI20171}. Thermal fluctuations are the random variations in the system's properties, for example, internal energy, that occur at thermal equilibrium. The energy temperature uncertainty relation links these fluctuations to the signal-to-noise ratio, quantum Fisher information~\cite{miller2018energy}, and provides a fundamental limit on the accuracy of quantum thermometers~\cite{jevtic_2015, Paris_2016}. Furthermore, in strongly coupled systems, thermal fluctuations become sensitive to quantum coherence and quantum correlations, necessitating a fresh exploration of classical thermodynamic relations in the quantum domain~\cite{Mukamel_2009, Brandes_2017}.

In recent times, efforts have been made to bridge the gap between quantum thermodynamics and metrology ~\cite{chu2022thermodynamic, cavina2018bridging, miller2018energy}. To this end, an important development in this field is a bound on signal-to-noise ratio for estimation of temperature ~\cite{miller2018energy}. This involves tools of quantum metrology ~\cite{giovannetti2011advances,paris2009quantum} in the form of the quantum Cramér-Rao inequality for an unbiased estimation of the temperature of the system. As is well known, the quantum Cramér-Rao inequality is bounded by the Fisher information ~\cite{liu2020quantum,helstrom1967minimum}. The other ingredient in this bound is the modified fluctuation-dissipation relation involving the specific heat capacity of the system as well as the Wigner-Yanase-Dyson skew information and another term that is pertinent to computing the system's internal energy ~\cite{miller2018energy, miller2018hamiltonian}.

In this article, we discuss the quantum thermodynamics of the following open quantum systems models: (i) a two-qubit model interacting with a thermal bath and (ii) the Jaynes-Cummings Model without the rotating wave approximation, using the Hamiltonian of Mean Force across all coupling regimes. Dynamics of two-qubit systems interacting with a squeezed thermal bath have revealed significant insights into entanglement dynamics, coherence, and quantum correlations~\cite{banerjee2007dynamics,banerjee2010entanglement,banerjee2010dynamics}. The Jaynes-Cummings model ~\cite{jaynes1963comparison,larson2021jaynes}, which describes a single two-level system interacting with a single-mode resonator, has seen significant progress in improving quantum measurement techniques, specifically in quantum sensing and metrology, and also in understanding the interplay between coherence and quantum thermal states~\cite{deffner2013quantum,casanova2010deep,xu2018non,messinger2020coherence}. We aim to understand how altering the bath temperature influences the thermal equilibrium characteristics of these models, and additionally, how thermal fluctuations affect temperature estimation through the signal-to-noise ratio.

The structure of the paper is as follows. The preliminary section (Sec.~\ref{sec_prelim}) includes an introduction to the Hamiltonian of Mean Force, certain thermodynamical potentials, thermodynamic quantifiers like ergotropy and entropy production, and a brief explanation of quantum information theoretic quantities such as WYD skew-information and Quantum Fisher information. In Sec.~\ref{Sec_III}, we briefly introduce the models under study and discuss their quantum thermodynamic properties using the Hamiltonian of Mean Force. This is followed by the conclusion in Sec.~\ref{conclusions}.

\section{Preliminaries}\label{sec_prelim}
This section is divided into two categories. The first category focuses on quantum information-theoretic quantities, where we define quantum uncertainty, skew information, and Fisher information. The second category pertains to quantum statistical mechanics, introducing the Hamiltonian of Mean Force and thermodynamic potentials such as entropy and specific heat capacity. Additionally, we provide a brief introduction to entropy production and ergotropy.

\subsection{Quantum Information-Theoretic Quantities}
\subsubsection{Quantum Uncertainty and Wigner-Yanase-Dyson Skew Information}
Skew information~\cite{wigner1963information} quantifies the quantum uncertainty in a state relative to an observable. 
It measures the non-commutativity between the state and the observable and has been extensively studied and generalized. 
The concept of skew information was extended through the definition \( Q_{\eta} (\rho, Y)= -\frac{1}{2}\Tr([\rho^{\eta},Y][\rho^{1-\eta},Y]) \), commonly referred to as WYD entropy~\cite{lieb1973convex}, where \( 0<\eta<1 \). When \(\eta = 1/2\), it simplifies to the well-known skew information. This generalized form can be expressed as:
\begin{equation}
    Q_{\eta} (\rho, Y)= \Tr(\rho Y^{2})- \Tr(\rho^{\eta} Y \rho^{1-\eta} Y).
\end{equation}
In comparison, the usual variance of an observable \(Y\) for the quantum state \(\rho\) is given by~\cite{li2011averaged},
\begin{equation}
    \text{Var}(\rho, Y) = \Tr(\rho Y^{2}) - (\Tr(\rho Y))^2= Q[\hat{\rho}, \hat{Y}]+ K[\hat{\rho}, \hat{Y}].
\end{equation}
The variance of the observable \(Y\) is influenced by both quantum and classical uncertainties inherent in the system. The measures of quantum and classical uncertainties, \( Q[\hat{\rho}, \hat{Y}] \) and \( K[\hat{\rho}, \hat{Y}] \), respectively, are defined as follows~\cite{miller2018energy}
\begin{equation}
    Q[\hat{\rho}, \hat{Y}] = \int_{0}^{1} Q_{a}[\hat{\rho}, \hat{Y}] \, da,
    \label{eq_quantum_uncertainty}
\end{equation}
where \( Q_{a}[\hat{\rho}, \hat{Y}] = - \frac{1}{2} \Tr\left( [\hat{Y}, \hat{\rho}^{a}] [\hat{Y}, \hat{\rho}^{1-a}] \right) \) with \( a \in (0,1) \). This measure \( Q[\hat{\rho}, \hat{Y}] \) quantifies the quantum uncertainty associated with the observable \( \hat{Y} \) in the state \( \hat{\rho} \). The non-commutativity of \( \hat{\rho} \) and \( \hat{Y} \) captures the essence of quantum fluctuations.
Similarly,
\begin{equation}
    K[\hat{\rho}, \hat{Y}] := \int_{0}^{1} K_{a}[\hat{\rho}, \hat{Y}] \, da,
    \label{CU}
\end{equation}
where \( K_{a}[\hat{\rho}, \hat{Y}] := \Tr\left( \hat{\rho}^{a} \delta \hat{Y} \hat{\rho}^{1-a} \delta \hat{Y} \right) \) with \( a \in (0,1) \). Here, \( \delta \hat{Y} = \hat{Y} - \langle \hat{Y} \rangle \) represents the deviation of \( \hat{Y} \) from its expectation value. The measure \( K[\hat{\rho}, \hat{Y}] \) captures the classical uncertainty, reflecting the statistical spread of outcomes due to classical probabilities. The parameter \( a \) indicates that there is no unique way of decomposing the variance into quantum and classical components. These quantum and classical uncertainty parameters play a very important role in the thermodynamics of quantum systems, which is explored in the later sections.

\subsubsection{Quantum Fisher Information}
Quantum Fisher Information (QFI) is a crucial concept in quantum metrology and information theory, representing the sensitivity of a quantum state to changes in a parameter. It is a fundamental tool for determining the ultimate precision limits in parameter estimation tasks, such as those encountered in quantum thermometry~\cite{paris2009quantum,wu2018local, chapeau2015optimized}. 
This quantity is pivotal as it establishes the quantum Cramér-Rao bound (QCRB), expressed as
\begin{equation}
\Delta \theta \ge \frac{1}{\sqrt{N F(\theta)}},
\label{cramerao}
\end{equation}
where \(N\) is the number of measurements (for estimation of $\theta$), dictating that the variance of any unbiased estimator \(\theta\) is inversely proportional to the QFI.\\

It has been recently proved ~\cite{miller2018energy} that for an exponential state $\hat{\rho}_{\theta}= \exp(-\hat{\Phi}_{\theta})/\mathcal{Z}_{\theta}$ the QFI concerning the parameter $\theta$ is bounded by,
\begin{equation}
    F(\theta) \leq K[\hat{\rho}_{\theta}, \hat{Y}_{\theta}],
    \label{fisherbound}
\end{equation} where, $\hat{Y}_{\theta}$ is the thermodynamic observable defined as $\hat{Y}_{\theta}=\partial_{\theta} \hat{\Phi}_{\theta}$ ,and $K[\hat{\rho}_{\theta}, \hat{Y}_{\theta}]$ is the classical uncertainty defined in Eq.~\eqref{CU}.
The spectral decomposition of the state $\hat{\rho}_{\theta}= \exp(-\hat{\Phi}_{\theta})/\mathcal{Z}_{\theta}$ is given by $\hat{\rho}_{\theta}= \sum_{n} p_{n} \big|\psi_{n}\rangle \langle \psi_{n} \big|$ where the eigenstates satisfy 
$\hat{\Phi}_{\theta} \big| \psi_{n}\big \rangle  = \lambda_{n}\big| \psi_{n} \big \rangle $. From this spectral decomposition, the QFI can be expressed as ~\cite{miller2018energy},
\begin{equation}
\label{qfi}
    F(\theta) = 2 \sum_{n,m} \frac{\left| \langle \psi_{n} \big| \partial_{\theta} \hat{\rho}_{\theta} \big| \psi_{m} \rangle \right|^2} {p_{n} + p_{m}}.
\end{equation} 

For a quantum thermal state of the form $\rho = e^{-\beta H }/{\rm Tr}\left[e^{-\beta H}\right]$ (where $\beta = 1/k_B T$ is the inverse temperature and $H$ is the Hamiltonian of the system), the Fisher Information for the temperature is proportional to the specific heat capacity of the system as $F(T)= C_{S}/T^2$ ~\cite{liu2020quantum}.

\subsection{Quantum Statistical Mechanics Related Quantities}
\subsubsection{Hamiltonian of Mean Force}
A quantum system $\mathcal{S}$ interacting with a thermal bath $\mathcal{B} $ is described by the Hamiltonian,
\begin{equation}
    \mathcal{\hat{H}_{T}} = \hat{H}_S \otimes \hat{\mathbb{I}}_B  + \hat{\mathbb{I}}_S \otimes  \hat{H}_B + \hat{\mathcal{V}}_{SB},
    \label{tothamil}
\end{equation}
where $\hat{H}_S$ and $ \hat{H}_B$ are the Hamiltonian of system and bath respectively, while $\hat{\mathcal{V}}_ {SB}$ is an interaction term of arbitrary strength. Here, we will consider situations where the environment is huge compared to the system, i.e., the operator norms fulfill $||\hat{H}_B|| \gg || \hat{H}_S||$, $||\hat{\mathcal{V}}_ {SB}||$.  
HMF takes into account the average effects of the environment on the subsystem.
The global equilibrium state of the system and the reservoir at temperature $T$ is denoted by $\hat{\zeta}_{SB}$, which is of Gibbs form,
\begin{equation}
    \hat{\zeta}_{SB} := \frac{\exp(-\beta \mathcal{\hat{H}_{T}})}{\mathcal{Z}_{SB}},
\end{equation}
where, $\mathcal{Z}_{SB} = \Tr_{SB} \left[\exp(-\beta \mathcal{\hat{H}_{T}})\right]$
is the partition function for $SB$. Also, $\beta = (k_{B}T)^{-1}$, the Boltzmann constant $k_{B}$ is set to unity throughout. The interaction term causes the reduced state of the system $S$ $\hat{\zeta}_{S}(T) = {\rm Tr}_{B} \left[\hat{\zeta}_{SB}(T)\right] $ to deviate from the thermal state unless the coupling is very weak (the magnitude of interaction term $\hat{\mathcal{V}}_{SB}$ is much lesser than the magnitude of the bare system Hamiltonian $H_S$). Thus, for the purpose of calculating internal energy in the strong coupling regime, the system's bare Hamiltonian would not be useful. 
This issue can be resolved by rewriting the state of the system $S$ as an effective Gibbs state $\hat{\zeta}_{S}(T) = \frac{\exp\left[-\beta \mathcal{\hat{H}}_{S}^* (T)\right]}{\mathcal{Z}_{S}^*}$, where the partition function for the system S can be expressed as the ratio 
\begin{equation}
    \mathcal{Z}_S^{*} = \mathcal{Z}_{SB}/\mathcal{Z}_{B},
\end{equation} where $\mathcal{Z}_{SB}= \Tr_{SB}[e^{-\beta \mathcal{\hat{H_{T}}}]}$ and $\mathcal{Z}_{B}= \Tr_{B}[e^{-\beta \hat{H}_{B} }]$ and,
\begin{equation}
    \mathcal{\hat{H}_{S}}^* (T) := - \frac{1}{\beta} \ln \left( \frac{ {\rm Tr}_{B} \left[\exp(-\beta \mathcal{\hat{H_{T}}}) \right]}{ {\rm Tr}_{B} \left[\exp(-\beta \hat{H_{B}})\right]}\right),
    \label{eq_Hamiltonian_of_mean_force}
\end{equation}
is the Hamiltonian of Mean Force ~\cite{kirkwood1935statistical,timofeev2022hamiltonian,miller2018hamiltonian}. 
This operator can be interpreted as an effective Hamiltonian describing $S$, and unlike the bare Hamiltonian $\hat{H}_{S}$, it implicitly depends upon both the temperature $T$ and the interaction $\hat {\mathcal{V}}_{SB}$. In the weak coupling regime, this reduces to the bare system Hamiltonian. HMF is essential for understanding thermodynamic properties such as free energy, entropy, and heat capacities in systems where direct interactions with an environment cannot be ignored. It allows for the formulation of generalized thermodynamic potentials and facilitates the study of equilibrium and non-equilibrium processes, providing insights into phenomena such as quantum dissipation and decoherence.

\subsubsection{Thermodynamic Potentials}
The partition function in statistical mechanics contains information regarding the distribution of occupation probabilities among the various microstates of the system $S$. It enables the calculation of important thermodynamic potentials like free energy, internal energy, and entropy.
The internal energy of the system $S$ can  be written as~\cite{miller2018hamiltonian},
\begin{equation}
    U_{S}(T) = U_{SB}(T) - U_{B}^{'}(T).
\end{equation}
This $U_{S}(T)$ is simply the difference between the total energy $U_{SB}(T)= -\partial_{\beta} \ln{\mathcal{Z}_{SB}}$ and the energy of the reservoir, $U_{B}^{'}(T)= -\partial_{\beta} \ln{\mathcal{Z}_{B}}$.
$U_{S}(T)$ can be expressed as an expectation value~\cite{seifert2016first,miller2018hamiltonian}, 
\begin{align}
U_{S}(T) &= \big \langle\hat{E}_S^*(T) \big\rangle = {\rm Tr} \big[ \hat{E}_S^*(T). \hat{\zeta}_S \big] \nonumber \\
&= {\rm Tr} \big[\partial_{\beta} ( \beta \mathcal{\hat{H}_{S}^* (T)} ). \hat{\zeta}_S \big] ,
\label{internalenergy}
\end{align}
of the observable,
\begin{equation}
 \hat{E}_S^*(T) = \partial_{\beta} [ \beta \mathcal{\hat{H}_{S}}^* (T) ] 
    = \hat{H}_{S}+  \partial_{\beta}[\beta (\mathcal{\hat{H}}_{S}^* (T)- \hat{H}_{S})].
    \label{Estar}
\end{equation}
This $\hat{E}_S^*(T)$ can be interpreted as the effective energy operator describing the system whose eigenstates are referred to as the system's energy eigenstates.  
With the inclusion of this operator, the fluctuations in the internal energy are considered as:
\begin{equation}
  \Delta U_{S}= \sqrt{{\rm Var}[\hat{\zeta}_S, \hat{E}_S^*]}.
  \label{eq_delta_US}
\end{equation}
It is significant to remember that $\hat{E}_S^*(T)$ depends on both the temperature $T$ and the system-environment coupling $\hat{\mathcal{V}}_ {SB}$. This $\hat{E}_S^*(T)$ differs from both the bare system Hamiltonian $\hat{H}_{S}$ and HMF $\mathcal{H}_S^*(T)$~\cite{miller2018energy}. From Eq.~\eqref{Estar} we can observe that only in those cases where $\mathcal{H}_S^*(T)$ becomes temperature independent will it become equal to $\hat{E}_S^*(T)$. 
The entropy of the system is calculated as
\begin{equation}
    \mathcal{S}_{S} = - {\rm Tr}[ (\ln  \hat{\zeta}_S) . \hat{\zeta}_S] + \beta^2 {\rm Tr}[ (\partial_{\beta}\mathcal{H}_S^*).\hat{\zeta}_S  ].
    \label{entropy}
\end{equation}
We recognize the first contribution as the von Neumann entropy of the system, while the second term reflects the temperature dependence of HMF. This indicates that when interactions are involved, the thermodynamic entropy generally does not correspond to the information present in the system's equilibrium state.

In the weak-coupling limit, it is well established that the specific heat capacity \(C_{S}\) is proportional to the variance in internal energy, expressed as ~\cite{miller2018hamiltonian}
\begin{equation}
C_{S} = \beta^2 \Delta U_{S}^2.
\end{equation}
This is a standard result of the fluctuation-dissipation relation, which connects the rate of change of the system's energy with temperature to the energy fluctuations at equilibrium. 
Recent research has indicated that for open quantum systems of the form given in Eq.~\eqref{tothamil}, the heat capacity can become negative at low temperatures, suggesting that it is not generally proportional to a positive variance ~\cite{binder2018thermodynamics,campisi2009fluctuation,hanggi2008finite,ingold2009specific}. Further, in the presence of strong coupling, the fluctuation-dissipation relation (FDR) includes two additional contributions~\cite{miller2018energy,miller2018hamiltonian},
\begin{equation}
C_{S}(\beta) = \beta^2 \Delta U_{S}^2 - \beta^2 Q[\hat{\zeta}_{S}, \hat{E}_{S}^*] - \frac{1}{\beta^2} \langle \partial_{\beta} \hat{E}_{S}^* \rangle,
\label{specific heat}
\end{equation}
where all the terms are as defined above.
This equation implies that \(C_{S}(\beta)\) can be less than \(\beta^2 \Delta U_{S}^2\) and even negative.
The first correction term arises from quantum fluctuations in energy, represented by the average WYD information related to the observable \(\hat{E}_{S}^*\). The second correction term \(\langle \partial_{\beta} \hat{E}_{S}^* \rangle\) accounts for the temperature dependence of \(\hat{E}_{S}^*\) and remains in the classical limit. In the weak coupling limit, where \(\hat{E}_{S}^* = \mathcal{\hat{H}}_{S}^*\), these correction terms disappear, making the heat capacity proportional to the energy variance.
When considering temperature as the unbiased parameter \(\theta\), the Cramer-Rao bound inequality in Eq.~\eqref{cramerao} can be written as \(\Delta \beta \ge \frac{1}{\sqrt{F(\beta)}}\) for a single measurement \(N=1\). Thus, Eq.~\eqref{fisherbound} can be reformulated as \(F(\beta) \leq K[\hat{\zeta}_{S}, \hat{E}_{S}^*]\). From this, a modified energy-temperature uncertainty relation was obtained as~\cite{miller2018energy}
\begin{equation}
\Delta \beta \geq \frac{1}{\sqrt{\Delta U_{S}^2 - Q[\hat{\zeta}_{S}, \hat{E}_{S}^*]}} .
\label{thermouncer}
\end{equation}
Comparing Eqs.~\eqref{specific heat} and \eqref{thermouncer}, the uncertainty in the temperature in the strong coupling regime can be expressed as
\begin{equation}
\Delta T \geq \frac{T}{\sqrt{C_{S}(T) - \langle \partial_{T} \hat{E}_{S}^* \rangle}} .
\label{deltaT}
\end{equation}
This expression sets a lower bound on the variance \(\Delta T\) (where a measurement is performed for estimating $T$). In the scenario of strong coupling, the best possible signal-to-noise ratio \(\frac{T}{\Delta T}\) for determining the temperature of the system \(S\) is constrained by both the specific heat capacity and an additional dissipation term \(\langle \partial_T \hat{E}^*_S \rangle\), which can be negative or positive. In the case of weak coupling, this additional dissipation term is zero. Therefore, in that case, the optimal signal-to-noise ratio \(\frac{T}{\Delta T}\) depends solely on the specific heat capacity, indicating that accurate temperature measurement requires a large heat capacity.
 
\subsection{Quantum thermodynamic quantifiers}
Here, we briefly discuss quantum thermodynamic quantifiers, particularly ergotropy and entropy production. 
\subsubsection{Ergotropy}
Ergotropy is the maximum work that can be extracted from a quantum system~\cite{PhysRevLett.85.1799,allahverdyan2004maximal}. Consider a quantum state $\rho$, whose spectral decomposition is given by 
\begin{align}
    \rho = \sum_i r_i\ket{r_i}\bra{r_i}, 
\end{align}
where $r_i$'s (in the order $r_1>r_2>r_3>...$) are the eigenvalues of $\rho$, and the spectral decomposition of the system's Hamiltonian $H_S$ is given by
\begin{align}
    H_S = \sum_j e_j\ket{e_j}\bra{e_j},
\end{align}
where $e_j$ (in the order $e_1<e_2<e_3...$) are eigenvalues of the system's Hamiltonian. The ergotropy for this system is given by 
\begin{align}
    \mathcal{W}[\rho(t)] = \sum_{j, i} r_ie_j\left(|\bra{r_i}\ket{e_j}|^2 - \delta_{ij}\right). 
\end{align}
In the case of open quantum systems, the ergotropy is calculated in a manner such that the state of the system, after partially tracing the bath out, is fed to the work extraction protocol involving unitary dynamics, quantified by ergotropy, {\it cf.}~\cite{Tiwari_Bhanja_2024, Tiwari2023}. Thus, the ergotropy provides the maximum extractable work from the state of the system after it is partially traced from the bath.

\subsubsection{Entropy production}
Irreversibility emerges in the system due to the following factors: (a) discarding the information locally contained in the state of the bath and (b) discarding the non-local information shared between the system and the bath, which is quantified using entropy production.
Consider a system bath ($S-B$) evolution of an arbitrary initial state $\rho_{SB}(0)$ dictated by the unitary operator $U$, such that $\rho_{SB}(t) = U\rho_{SB}(t)U^\dagger$. Entropy production~\cite{esposito2010entropy, landi} is given by 
\begin{align}
    \Sigma = \mathcal{I}(S:B) + S\left[\rho_{B}(t)||\rho_B(0)\right],
\end{align}
where $\mathcal{I}(S:B)$ is the mutual information given by the von Neumann entropies ($S(\rho) = -{\rm Tr}(\rho\ln\rho)$) of the subsystems $S$ and $B$ and the joint system $SB$, that is, 
\begin{align}
    \mathcal{I}(S:B) = S[\rho_S(t)] + S[\rho_B(t)] - S[\rho_{SB}(t)].
\end{align}
$S(\rho||\sigma) = {\rm Tr} \left(\rho\ln\rho - \rho\ln\sigma\right)$ is the quantum relative entropy. The above relation can further be simplified as
\begin{align}
    \Sigma = S\left[\rho_{SB}(t)||\rho_S(t)\otimes\rho_B(0)\right],
    \label{eq_entropy_production}
\end{align}
which is the Kullback-Leibler divergence between the joint system-bath evolved state and the state $\rho_S(t)\otimes\rho_B(0)$, where $\rho_S(t) = {\rm Tr}_B\left(U\rho_{SB}(t)U^\dagger\right)$. 

\section{Illustrative Open Quantum System Models}\label{Sec_III}
Here, we briefly discuss the models under consideration in this work. This is followed by a study of their corresponding quantum thermodynamic properties. The models studied in this work provide a thermodynamic perspective on the role of system-environment interactions. By examining both weak and strong coupling regimes, we uncover how non-Markovian memory effects prominent in the strong coupling limit influence thermodynamic quantities such as specific heat, ergotropy, and entropy production.

\subsection{Two-qubit system interacting with a thermal bath}
We start with a two-qubit system interacting with a thermal bath via a dissipative interaction. 
The dissipative interaction between 2-qubits (a two-level atomic system) and the bath (represented as a three-dimensional electromagnetic field (EMF)) via the dipole interaction ~\cite{ficek2002entangled} is described by the Hamiltonian,
\begin{align}
H &= H_S + H_B + H_{SB} \nonumber \\
                 &= \sum_{j=1}^2 \hbar \omega_j S_j^{z} \nonumber 
                 + \sum_{\vec{k}s} \hbar \omega_k \left( \alpha_{\vec{k}s}^\dagger \alpha_{\vec{k}s} + \frac{1}{2} \right) \nonumber \\
                 &\quad - i\hbar \sum_{\vec{ks}} \sum_{j=1}^{2} [ \vec{\mu}_j \cdot \vec{g}_{\vec{ks}} \left( \vec{r}_j \right)(S_{j}^{+} + S_{j}^{-}) \alpha_{\vec{k}s} - \text{h.c.} ]
\label{eq:total_hamiltonian}
\end{align}
The two-qubits are modeled as two-level systems with excited state $\big|e_{j} \rangle$, ground state $\big|g_{j} \rangle$, transition frequency $\omega_{j}$, and transition dipole moments $\Vec{\mu_{j}}$. We assume that the qubits are located at different atomic positions $\Vec{r_{j}}$. The transition dipole moments are dependent on $\Vec{r_{j}}$.
$S_{j}^+ = \ket{e_{j}} \bra{g_{j}}$ and $S_{j}^- = \ket{g_{j}}\bra{e_{j}}$ are the dipole raising and lowering operators, respectively, satisfying the well-known commutation and anti-commutation relations $[S_{j}^+, S_{k}^-] = 2S_{j}^z \delta_{jk}$; $[S_{j}^z, S_{k}^{\pm}]= \pm S_{j}^{\pm} \delta_{jk}$; $[S_{j}^+, S_{k}^-]_+= \delta_{jk}$ with $(S_{j})^2 \equiv 0$ and 
$S_{j}^{z} = \tfrac{1}{2} \left(\ket{e_{j}}  \bra{e_{j}} - \ket{g_{j}}\bra{g_{j}}\right)$
is the energy operator of the jth-atom. 
$\alpha_{\vec{k}s}^\dagger$, $ \alpha_{\vec{k}s}$ are the creation and annihilation operators of the field mode $\vec{k}s$ with wave vector $\vec{k}$, frequency $\omega_{k}$ and the polarization index $s$. The system-reservoir coupling constant is
\begin{equation}
    \Vec{g}_{\vec{k}s} (\vec{r}_j) = \left(\frac{\omega_{k} }{2\epsilon_0 \hbar V}\right)^{1/2} \Vec{e}_{\vec{k}s} e^{i \vec{k}.\vec{r}_{j}}.
\end{equation}
$V$ is the normalization volume, $\epsilon_0$ is the permittivity of free space, and $\Vec{e}_{\vec{k}s}$ is the unit polarization of the field. This equation implies that the system-reservoir coupling constant depends on the atomic position $\vec r_{j}$. This leads to two regimes: one where the inter-qubit spacing is less than the length scale set by the bath (via its resonant wavelength), that is, the collective regime, and the other where the inter-qubit spacing is comparable with the length scale, the independent regime~\cite{banerjee2010dynamics}. In this work, the collective regime is considered for further analysis. 

For simplification, we assume that both the qubits have the same transition frequency and the first qubit is at the origin, while the distance between them is $\xi$. Further, $\vec{k}$ is taken to be along $\vec{r}_{12}$. The dipole moments $\vec{\mu}_j$'s are taken to be equal for both qubits ( $\mu_1 = \mu_2 = \mu$) and along the direction of a polarization vector, such that $\vec{\mu}_j.\vec{e}_{\vec{k}s} = \mu$. Incorporating these quantities, the Hamiltonian in Eq.~\eqref{eq:total_hamiltonian} (for $\hbar = c = 1$) can be rewritten as
\begin{align}
    H &= \omega_0\left(S_1^z + S_2^z\right) + \sum_{k} \omega_k \left( \alpha_{k}^\dagger \alpha_{k} + \frac{1}{2} \right) \nonumber \\
    &-i \sum_k\lambda\sqrt{\omega_k}\left[\left(S_1^+ + S_1^-\right)\left(\alpha_k - \alpha_k^\dagger\right)\right.\nonumber \\
    &\left. +\left(S_2^+ + S_2^-\right)\left(\alpha_ke^{i\omega_k\xi} - \alpha_k^\dagger e^{-i\omega_k\xi}\right) \right],
    \label{eq_2_qubit_multimode_hamiltonian}
\end{align}
where $\lambda  = \left(\frac{\mu}{\sqrt{2\epsilon_0V}}\right)$. 
The reduced density matrix of the two-qubit system in the interaction picture and the standard Born-Markov rotating wave approximation (RWA) was obtained in~\cite{banerjee2010dynamics,ficek2002entangled} by assuming a separable initial system-bath state and taking a trace over the bath. 
When $\lambda \sim \omega_0$, the interaction strength between the system and the bath is comparable to the system's natural frequency $\omega_0$. This characterizes a strong coupling regime, wherein the system and the bath can efficiently exchange energy, resulting in significant mutual influence on each other's dynamics. 
Conversely, when $\lambda \ll \omega_0$, the interaction strength is considerably smaller than the system's natural frequency. This corresponds to a weak coupling regime, where the interaction between the system and the bath is minimal. Consequently, the system's dynamics are less perturbed by the bath, and energy exchange between the two is smaller.

\begin{figure}
    \centering
    \includegraphics[width=1\linewidth]{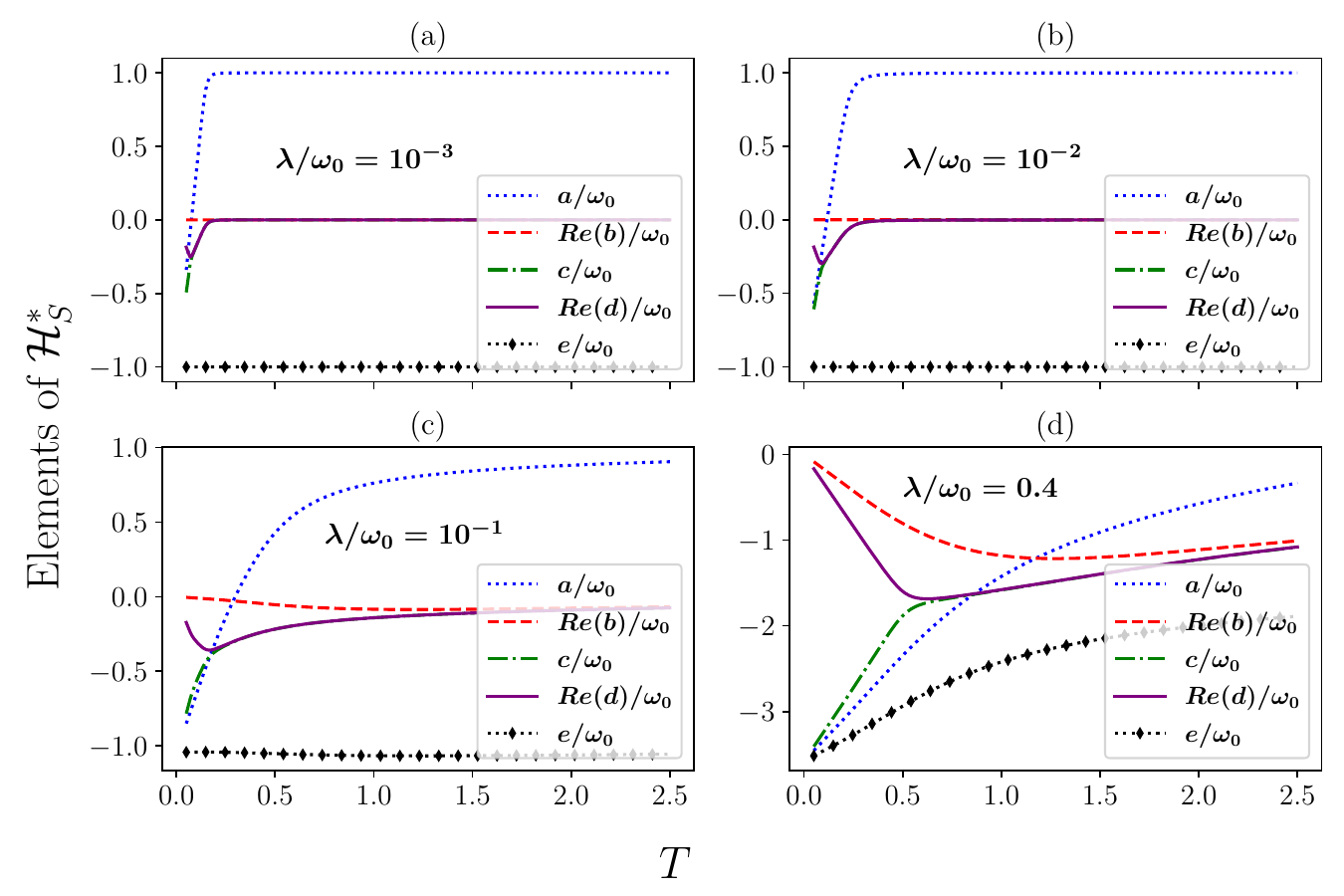}
    \caption{The variation of the elements of the Hamiltonian of Mean Force $\mathcal{H}^*_S$ for the two-qubit model as a function of the temperature for different coupling strengths $\lambda$ in the collective regime. Here, we have taken $\omega_0 = 2$, and $\xi = 0.05$.}
    \label{fig_2_qubit_hmf_elements}
\end{figure}

\begin{figure}
    \centering
    \includegraphics[width=1\columnwidth]{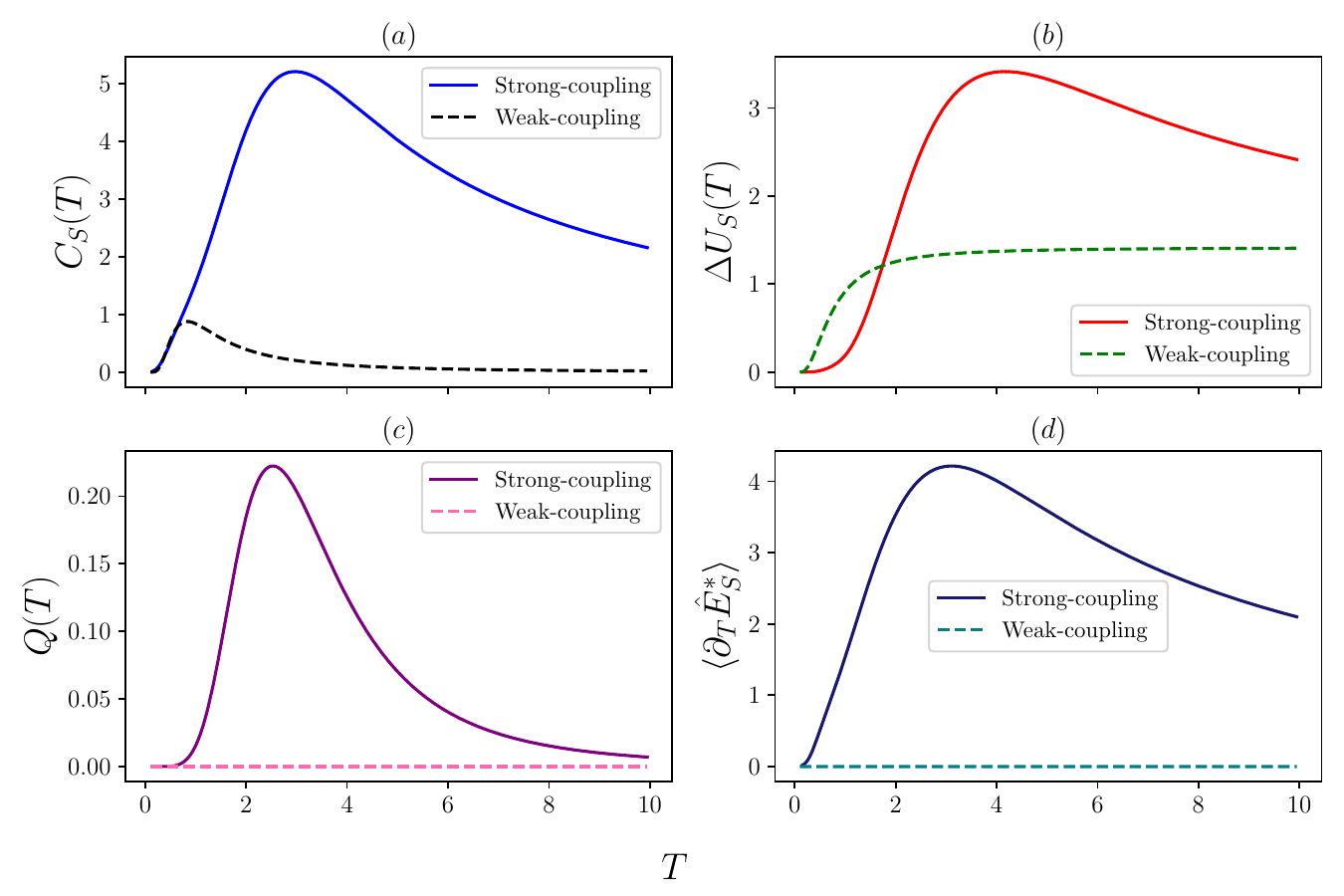}
    \caption{Variation of (a) specific heat capacity $C_S(T)$, (b) fluctuations in the internal energy $\Delta U_S(T)$, (c) quantum uncertainty $Q(T)$, and (d) $\langle\partial_T \hat E^*_S\rangle$ with temperature $T$. The parameters are $\omega_0 = 2.0, \xi = 0.05$. In strong coupling $\lambda/\omega_0 >0.1$, while in weak coupling limit, it is $\sim 10^{-3}$.}
    \label{fig_2_qubit_decoherence_model_heat_capacity}
\end{figure}

HMF for this model is calculated using Eqs. \eqref{eq_Hamiltonian_of_mean_force} and \eqref{eq_2_qubit_multimode_hamiltonian}, where $\mathcal{H}_\mathcal{T}$ is the total Hamiltonian of the system and bath $H$. To compute the HMF numerically, particularly in the strong-coupling regime, we have taken the number of modes of the resonator computationally possible with different $\omega_k$'s. 
The number of modes that can be taken this way is limited due to computational complexity, as described below. The bath is modeled as comprising $N$ discrete resonator modes. Each mode $k$ is described by a Fock space, which is truncated at a finite dimension $d(\mathcal{H}_{B, k})$ ($\mathcal{H}_{B, k}$ denotes Fock space of the $k$th resonator mode) when the elements of the HMF do not significantly change by increasing the Fock space. The total Hilbert space of the system plus bath is given by the tensor product $\mathcal{H}_{\rm tot} = \mathcal{H}_S\otimes \mathcal{H}_{B,1}\otimes \mathcal{H}_{B,2}\cdots\otimes\mathcal{H}_{B, N}$, where $\mathcal{H}_S$ is the system's Hilbert space. Consequently, the total dimension of the joint space scales as $d(\mathcal{H}_S)\times \prod_{k =1}^N d(\mathcal{H}_{B, k})$, leading to exponential growth in computational complexity with increasing $N$.
It has been shown that the computation of numerical HMF can be performed robustly by utilizing the time-evolving matrix product operator~\cite{PhysRevA.106.012204}. The numerical form of the HMF for this model is given by 
\begin{align}
    \mathcal{H}^*_S = \begin{pmatrix}
        a&0&0&b\\
        0&c&d&0\\
        0&d^*&c&0\\
        b^*&0&0&e
    \end{pmatrix},
    \label{hmf_form}
\end{align}
in the energy eigenbasis of the system's bare Hamiltonian (computational basis). Notice that the structure of the above Hamiltonian is an $X$-type matrix and is Hermitian. The elements $a, b, c, d$, and $e$ depend on the temperature $T$ and the coupling strength $\lambda$ between the system and the bath, as shown in Fig.~\ref{fig_2_qubit_hmf_elements}.
The two-qubit system's Hamiltonian is $H_S = \omega_0\left(\ket{0}\bra{0} - \ket{3}\bra{3}\right)$ in the computational basis. It is observed that in the weak coupling limit $\lambda/\omega_0 < 10^{-1}$ and for higher temperatures, the off-diagonal elements $b$ and $d$, and the element $c$ become zero. Further, the elements $a$ and $e$ become $\omega_0$ and $-\omega_0$, respectively. Therefore, in this limit, the HMF becomes equal to the bare Hamiltonian of the system $H_S$. In the strong coupling limit $\lambda/\omega_0\sim 0.4$, it can be observed that all the elements $a, b, c, d$ and $e$ are non-zero and are initially negative. At higher temperatures, the elements $c$ and the ${\rm Re}(d)$ are equal, and the element ${\rm Re}(b)$ is asymptotically equal to both. Interestingly, the effective Gibbs state based on HMF, Eq.~\eqref{hmf_form}, has coherent terms in the energy eigenbasis of the bare Hamiltonian (computational basis) in the strong coupling regime and at lower temperatures. This observation is consistent with~\cite{cresser2021weak}.  
The specific heat capacity and the corresponding variables in its expression are plotted in Fig.~\ref{fig_2_qubit_decoherence_model_heat_capacity}.
The specific heat capacity of a system quantifies the susceptibility of the system to change its energy with respect to variation in temperature. At low temperature, the system is mostly in its ground state, and there is very little energy exchange with the bath, resulting in small internal energy fluctuations and low specific heat. At high temperatures, the thermal energy becomes large, and all energy levels have a tendency to become equally populated. Due to this, the system is less susceptible to temperature changes, leading to a small heat capacity. Interestingly, at intermediate temperature, when the system's energy level gaps become comparable to the thermal energy, the specific heat capacity and internal energy fluctuations attain a maximum. This is more evident in the regime where the system is strongly coupled to the bath.
As can be observed, the quantum uncertainty $Q(T)$ and the quantity $\langle\partial_T \hat E^*_S\rangle$ are zero in the weak coupling regime but non-zero in the strong coupling regime. This is consistent with the generalized fluctuation-dissipation relation, Eq.~\eqref{specific heat}. The internal energy $\Delta U_S(T)$ increases and then saturates with an increase in temperature. In the strong coupling regime, $\Delta U_S(T)$ attains a maximum and then decreases with an increase in temperature. The pattern of the specific heat capacity $C_S(T)$ is similar in both the weak and strong coupling regimes. However, in the weak-coupling regime, the specific heat capacity remains relatively low and decreases gradually as temperature increases.
In the strong-coupling regime, the emergence of quantum uncertainty and significant corrections to the specific heat capacity indicate the presence of memory effects in the system-environment interaction. This reflects characteristic features of non-Markovian dynamics as captured through thermodynamic signatures. Such behavior is absent in the weak-coupling limit.

\begin{figure}
    \centering
    \includegraphics[width=1\linewidth]{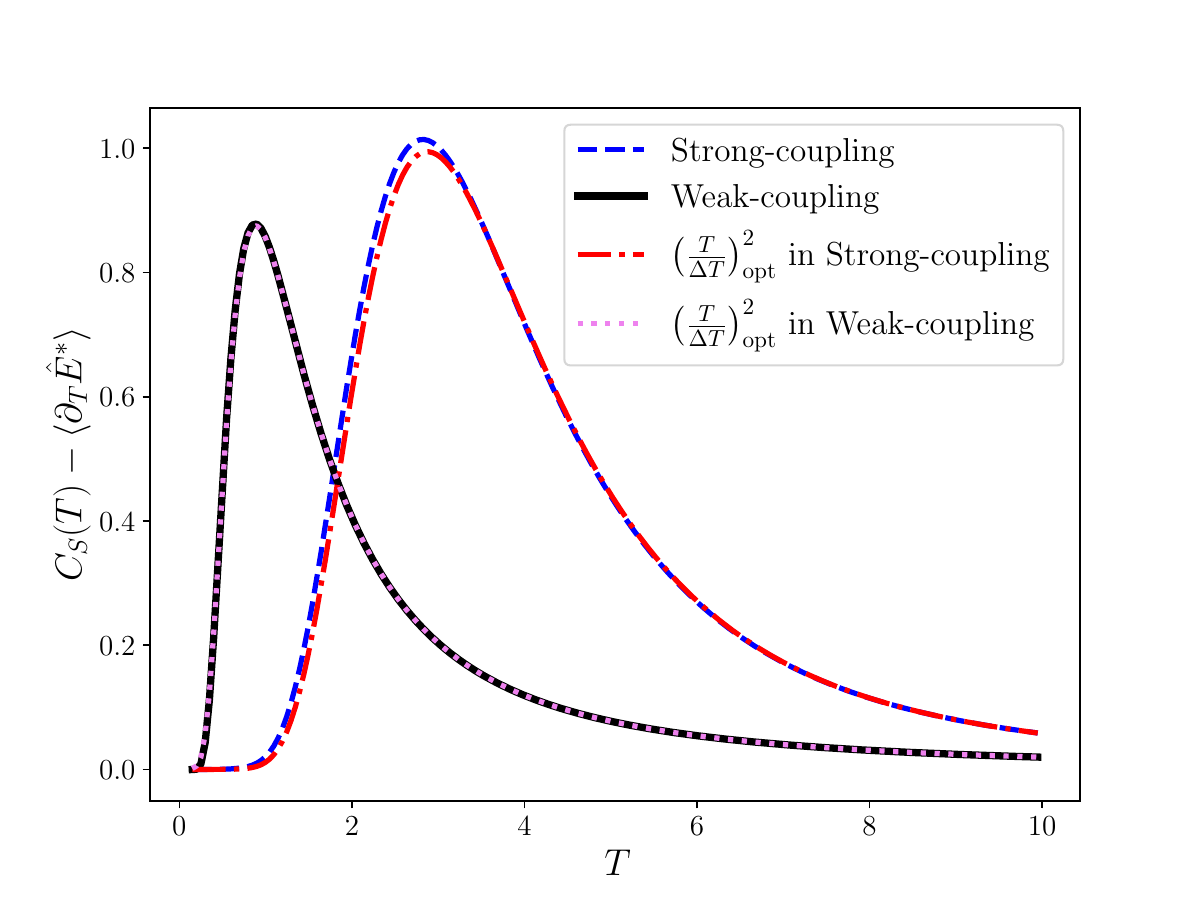}
    \caption{Variation of the upper bound $C_S(T) - \langle\partial_T \hat E^*\rangle$ on the signal-to-noise ratio $\left(\frac{T}{\Delta T_S}\right)^2$, and of $\left(\frac{T}{\Delta T_S}\right)^2_{\rm opt} = T^2F_S(T)$ as a function of temperature $T$. The parameters are $\omega_0 = 2.0, \xi = 0.05$. In strong coupling $\lambda/\omega_0 >0.1$, while in weak coupling limit, it is $\sim 10^{-3} \omega_0$.}
    \label{fig_upper_bound_2_qubit_model_1}
\end{figure}

An upper bound on the signal-to-noise ratio, obtained from the modified energy-temperature uncertainty relation, Eq.~\eqref{deltaT}, is depicted in Fig.~\ref{fig_upper_bound_2_qubit_model_1}. 
The bounds are well satisfied for both the weak and strong coupling regimes, with the weak coupling regime depicting the tighter bound, in fact, saturating the bound. This saturation of the bound can be linked with the structure of the HMF, Eq.~\eqref{hmf_form}, which becomes diagonal in the energy eigenbasis of the bare Hamiltonian in the weak coupling and high-temperature regime, and equal to the bare Hamiltonian of the system.

\begin{figure}
    \centering
    \includegraphics[width=1\columnwidth]{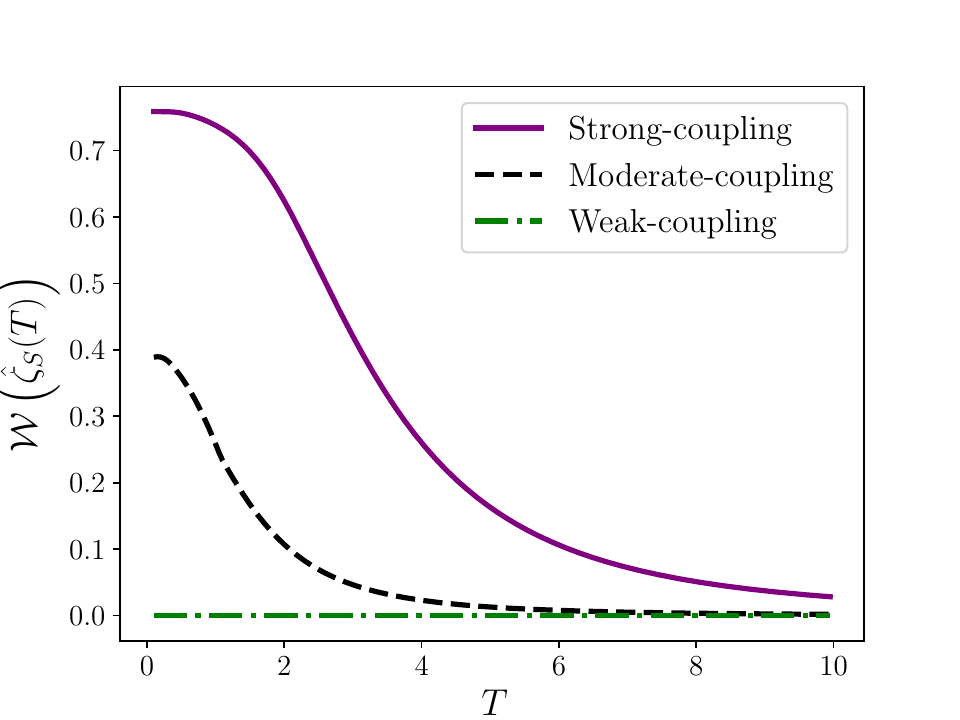}
    \caption{Variation of Ergotropy $\mathcal{W}\left(\hat \zeta_S\right)$ with Temperature $T$ in the strong, moderate, and weak coupling limit for a two-qubit squeezed thermal bath model. The parameters are chosen to be $\omega_0 = 2.0, \xi = 0.05$. In strong coupling $\lambda/\omega_0 >0.1$, moderate coupling is when the coupling constant is kept half of the strong-coupling constant value, while in the weak coupling limit, it is $\sim 10^{-3} \omega_0$.}
    \label{ergotropy in two-qubit}
\end{figure}

The ergotropy is calculated using the state $\zeta_S(T) = e^{-\beta \mathcal{H}^*_S}/Z^*$, and is plotted in Fig.~\ref{ergotropy in two-qubit}.
From this figure, one can infer that in the weak coupling regime, the HMF coincides with the bare Hamiltonian of the system, and the ergotropy becomes zero, as the state $\zeta_S(T)$ becomes the Gibbs state for the bare Hamiltonian, which is not the case in the moderate and strong coupling regimes. Further, ergotropy is composed of coherent and incoherent parts. It is observed that for the state $\zeta_S(T)$ of the two-qubit model, the incoherent contribution is zero. The non-zero ergotropy comes from the coherent part. This is due to the fact that the state $\zeta_S(T)$ has coherence terms present in it at stronger system-bath couplings, which are absent in the weak coupling regime. The appearance of non-zero ergotropy, arising from coherent components of the system state under strong coupling, highlights the re-emergence of usable quantum coherence, which is a hallmark of non-Markovian dynamics even in the presence of thermalization.

\begin{figure}[h]
    \centering
    \includegraphics[width=1\linewidth]{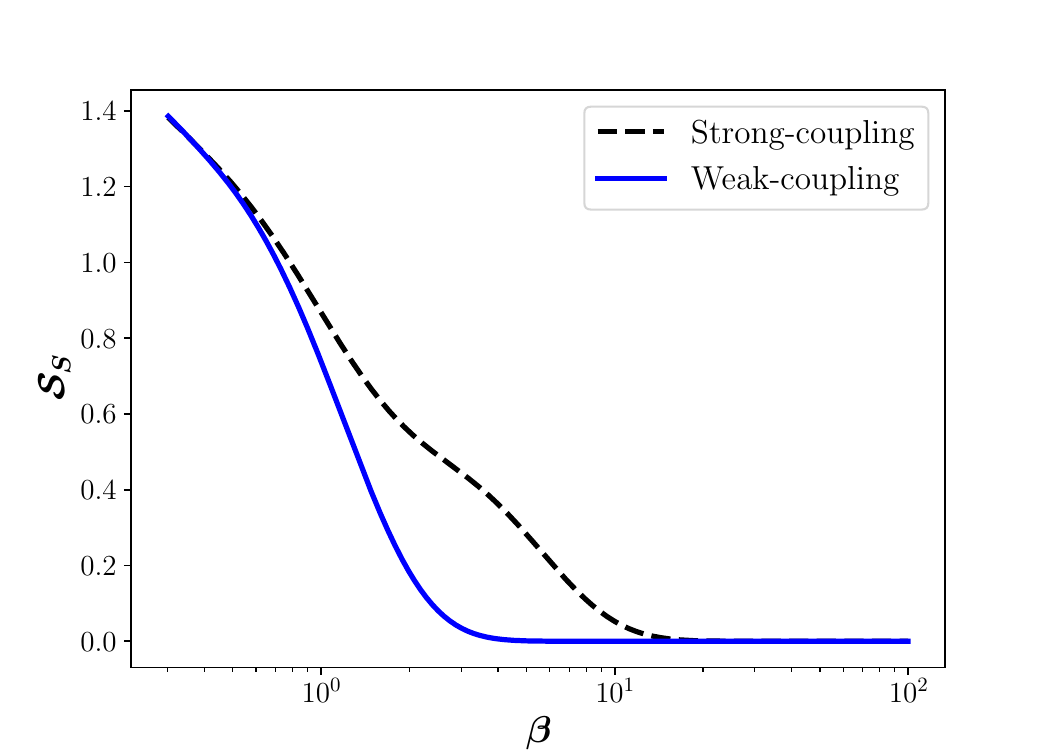}
    \caption{Plot for the entropy [Eq.~\eqref{entropy}] as a function of temperature for the two-qubit decoherence model. The parameters are chosen to be $\omega_0 = 2.0, \xi = 0.05$. In strong coupling $\lambda/\omega_0 >0.1$, while in weak coupling limit, it is $\sim 10^{-3} \omega_0$.}
    \label{fig_entropy_plot_2_qubit_model}
\end{figure}

To examine the third law of thermodynamics, we calculate the entropy (using HMF), which is plotted in Fig.~\ref{fig_entropy_plot_2_qubit_model}. 
It can be observed from the plot that for lower temperatures (higher $\beta$), the entropy tends to zero for both weak and strong system-bath couplings, consistent with the third law of thermodynamics. Further, at higher temperatures, that is, at lower $\beta$, the entropy is approximately the same in both weak and strong coupling regimes. The difference in the entropy for the weak and strong system bath couplings can be seen, in particular in the intermediate temperature regime, where entropy has higher values at relatively low temperatures in the strong coupling regime. 

Further, to study the second law of thermodynamics, we calculate the entropy production of the system for a single bath mode, Eq.~\eqref{eq_entropy_production}. To this end, we use the Bell state $\frac{1}{\sqrt{2}}\left(\ket{00} + \ket{11}\right)$ as the initial state of the two-qubit system, and $\rho_B(0) = \frac{e^{-\beta H_B}}{{\rm Tr}\left(e^{-\beta H_B}\right)}$ as the initial state of the bath. The dynamics of the total and the reduced systems are obtained using $U \left\{\rho_S(0) \otimes \rho_B(0)\right\} U^\dagger$ and ${\rm Tr}_B \left[U \left\{\rho_S(0) \otimes \rho_B(0)\right\} U^\dagger\right]$, respectively, where $U = e^{-iHt}$. 
\begin{figure}
    \centering
    \includegraphics[width=1\linewidth]{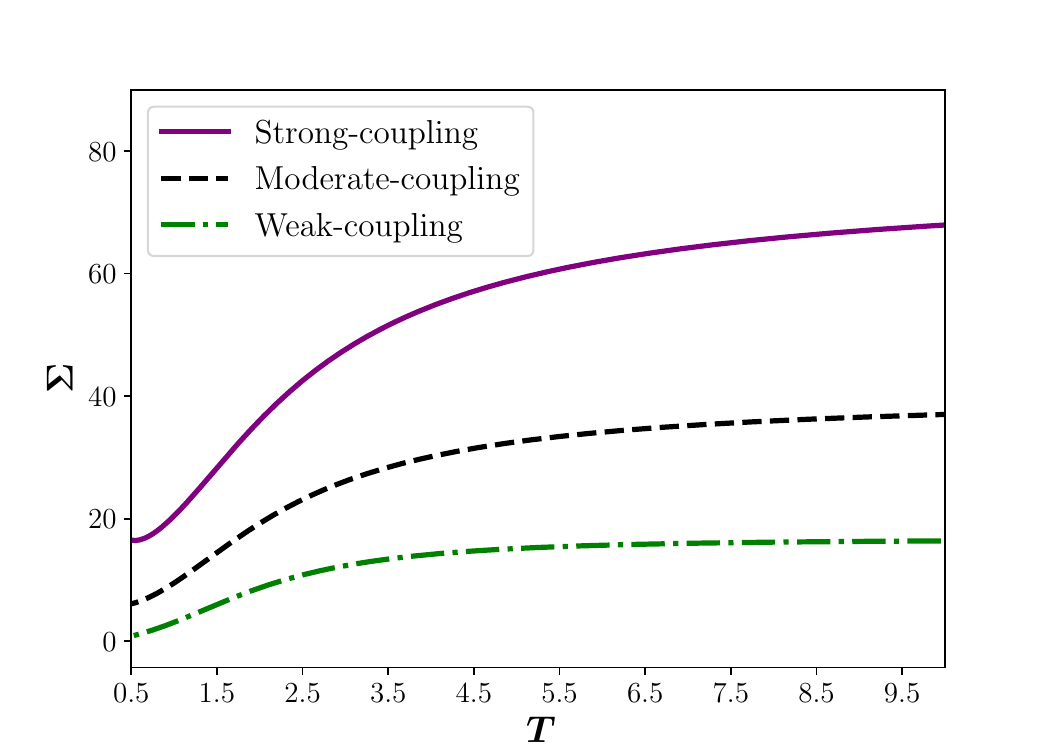}
    \caption{Variation of the entropy production $\Sigma$ as a function of temperature at time $t=1$ for the two-qubit model interacting with a single bath resonator. The strong, moderate, and weak couplings are taken as discussed in the above figures. The parameters are taken to be: $\omega_0 = 2.0, \omega = 1, \xi = 0.05$.}
    \label{fig_Entropy_production_2_qubit_model}
\end{figure}%
The variation of entropy production $\Sigma$ as a function of temperature, at time $t = 1$, is plotted in Fig.~\ref{fig_Entropy_production_2_qubit_model}. As can be observed from Fig.~\ref{fig_Entropy_production_2_qubit_model}, the entropy production is the highest in the case of strong system-bath coupling, which decreases as we decrease the coupling strength. The entropy production shows a saturating behavior as we increase the temperature. The entropy production remains non-negative for all values of temperature $T$, in accordance with the second law of quantum thermodynamics.

\subsection{A single two-level system interacting with a single-mode resonator}\label{sec_nmgad_model}
\begin{figure}
    \centering
    \includegraphics[width=1\linewidth]{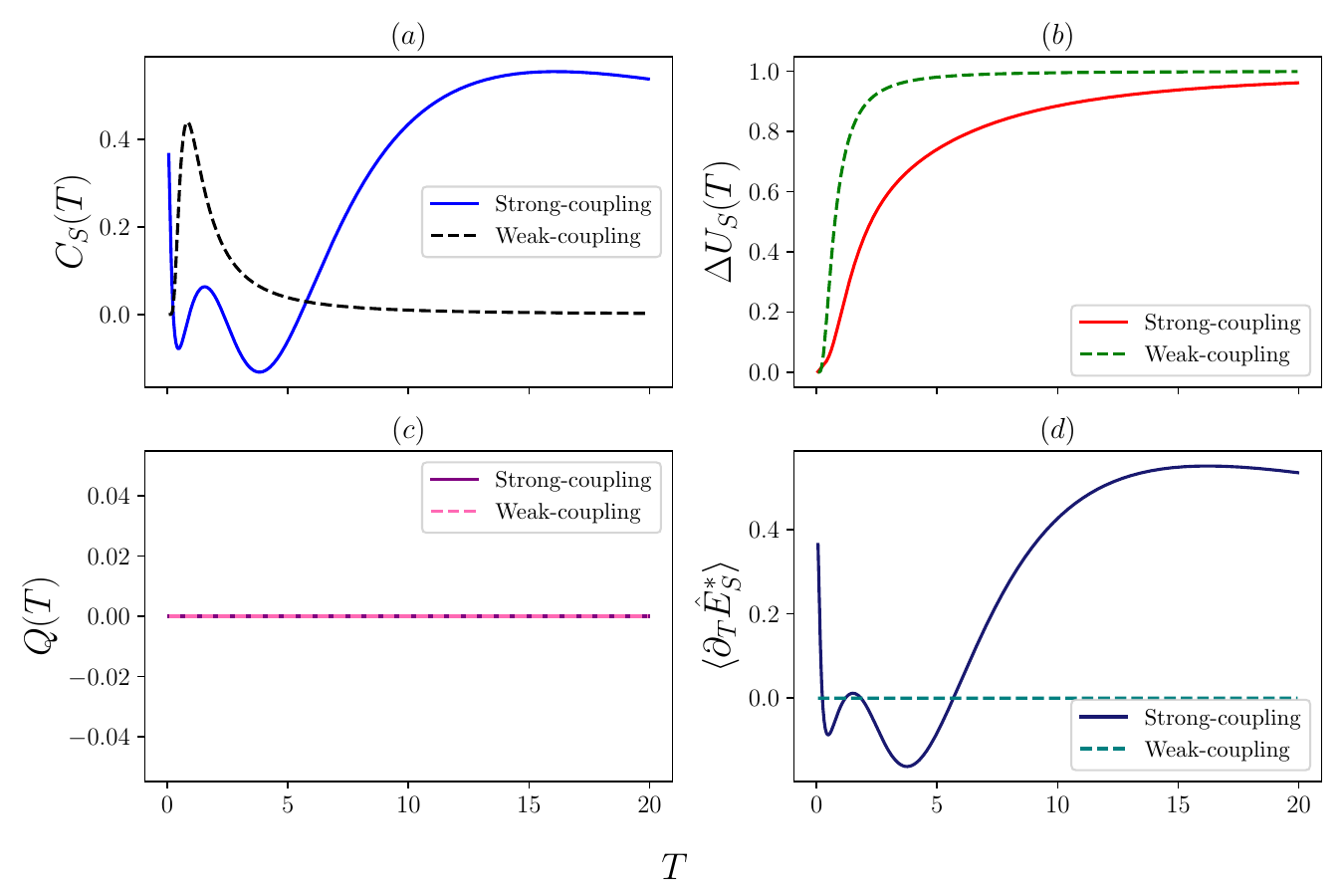}
    \caption{Variation of (a) specific heat capacity $C_S(T)$, (b) fluctuations in the internal energy $\Delta U_S(T)$, (c) quantum uncertainty $Q(T)$, and (d) $\langle\partial_T \hat E^*_S\rangle$ with temperature $T$ for the JC model. The parameters are: $\omega_0 = 2.0, \omega_c = 1.0$. The weak coupling limit corresponds to $\lambda \ll \omega_0$, and $\lambda\sim \omega_0$ corresponds to the strong coupling limit.}
    \label{fig_1_qubit_decoherence_model_heat_capacity}
\end{figure}

Here, we take an example of a two-level system interacting with a single-mode field, that is, the Jaynes-Cummings (JC) model without rotating wave approximation~\cite{larson2021jaynes, vacchini}. The Hamiltonian (for $\hbar=1$) of the total system is given by 
\begin{align}
    H &= H_S + H_B + H_{SB}\nonumber \\
    &= \frac{\omega_0}{2}\sigma^z + \omega_c \hat a^\dagger \hat a + \lambda \sigma^x\left(\hat a + \hat a^\dagger \right).
\end{align}
Diagonalizing the above Hamiltonian is highly non-trivial, and therefore, we partially trace the field and find the HMF numerically (here, we truncate the Fock space corresponding to the field mode at $n$ when the factors $p$ and $q$, see equation below, do not change significantly upon increasing the value of $n$, such that $|\left(\text{value of}~p, q~\text{at}~n\right) - \left(\text{value of}~p, q~\text{at}~n - 1\right)|<\epsilon$, where $\epsilon$ is of the order of $10^{-8}$) for this model using Eq.~\eqref{eq_Hamiltonian_of_mean_force}, and is given by
\begin{align}
  \mathcal{H}^*_{S} =  p\ket{0}\bra{0} + q\ket{1}\bra{1}, 
  \label{eq_single_qubit_hmf}
\end{align}
where $\ket{0} = \left(1~~0\right)^T$, and $\ket{1} = \left(0~~1\right)^T$. As can be seen from the above equation, $\mathcal{H}^*_S$ is diagonal in the energy eigenbasis of the system's bare Hamiltonian (computational basis), though not equal to $H_S$ in the strong coupling and low-temperature regime. It is, however, the same as the bare Hamiltonian in the weak coupling and high-temperature regime.  
\begin{figure}
    \centering
    \includegraphics[width=1\columnwidth]{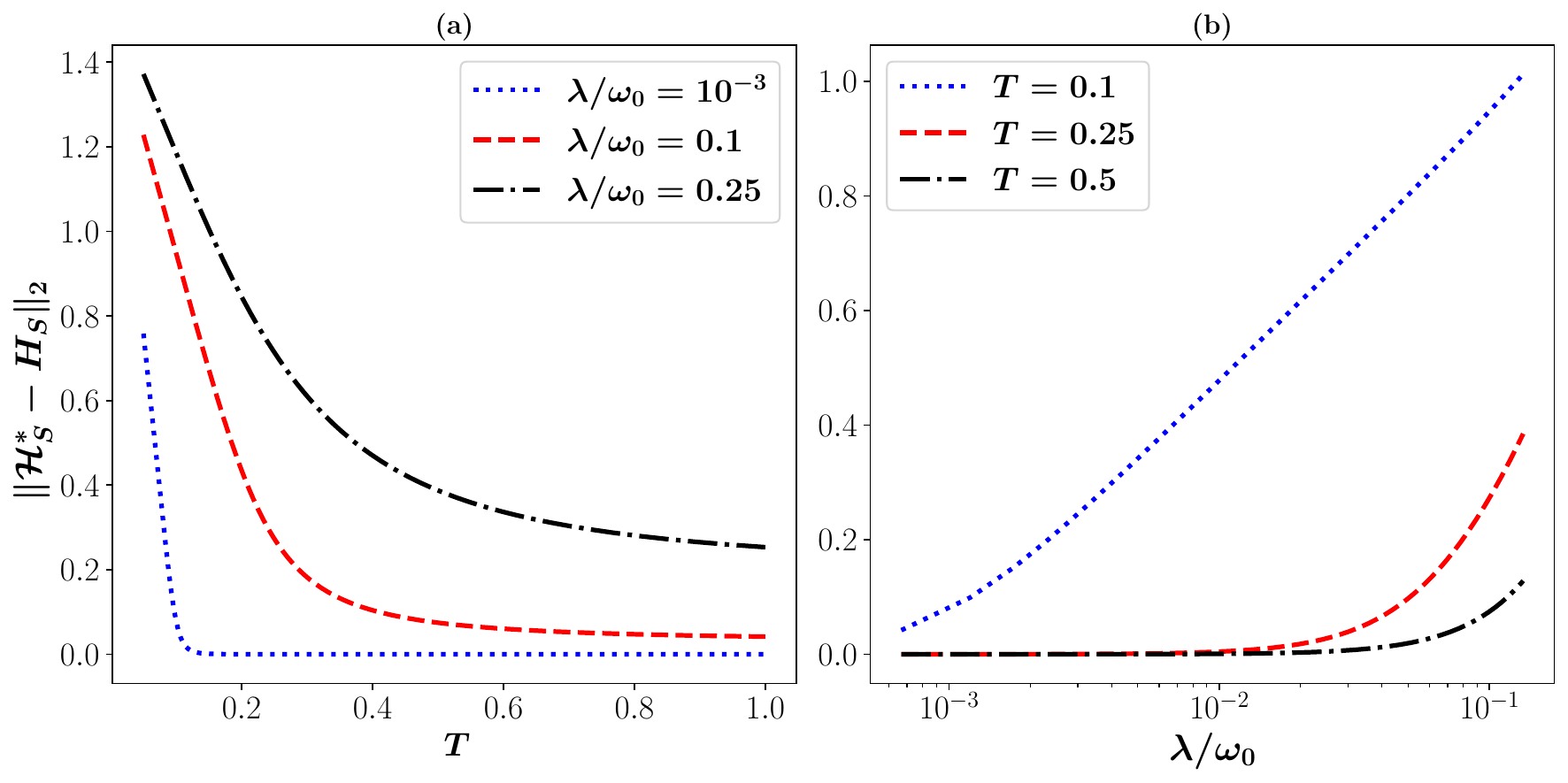}
    \caption{Variation of the Hilbert-Schmidt norm $\|\mathcal{H}_S^* - H_S\|_2$ as a function of (a) temperature $T$ and (b) coupling strength $\lambda$ for the JC model. In (a), the plots are for different coupling strengths, and in (b), the plots are for different temperatures. The parameters are: $\omega_0 = 1.5, \omega_c = 1.0$.}
    \label{fig_hs_norm_single_qubit_model}
\end{figure}
This behavior of the HMF can be observed in Fig.~\ref{fig_hs_norm_single_qubit_model}, where we plot the variation of the Hilbert-Schmidt norm $\|\mathcal{H}_S^* - H_S\|_2$ as a function of temperature and the system bath coupling strength. The norm of the difference between the HMF and the bare Hamiltonian of the system increases as we decrease the temperature and increase the coupling strength. At higher temperatures and lower coupling strengths, the HMF is equal to the bare Hamiltonian of the system. A similar behavior was observed for a spin-chain system in~\cite{PhysRevE.110.014111}. Further, the factors $p$ and $q$ depend on the temperature $T$ and the coupling strength between the system and the bath, and in the weak coupling and high-temperature limit $p\rightarrow\frac{\omega_0}{2}$ and $q\rightarrow\frac{-\omega_0}{2}$.
The effective Gibbs state $\hat \zeta_S (\beta)$ using the HMF for the single-qubit model, Eq.~\eqref{eq_single_qubit_hmf}, is given by
\begin{align}
    \hat \zeta_S(\beta) = \frac{e^{-\beta \mathcal{H}^*_S}}{{\rm Tr}\left[e^{-\beta \mathcal{H}^*_S}\right]} = \begin{pmatrix}
        \frac{1}{1 + e^{\beta(p-q)}} &0\\
        0&\frac{1}{1 + e^{\beta(q-p)}}
    \end{pmatrix},
\end{align}%
which becomes $e^{-\beta H_S}/{\rm Tr}(e^{-\beta H_S})$ in the weak coupling and high $T$ limit. Further, the term $\hat E^*_S$ is given by $\left(p + \beta \partial_\beta p\right)\ket{0}\bra{0} + \left(q + \beta \partial_\beta q\right)\ket{1}\bra{1}$. The specific heat capacity $C_S(T)$ and related quantities as a function of temperature are plotted in Fig.~\ref{fig_1_qubit_decoherence_model_heat_capacity}. 
It is observed that the quantum uncertainty $Q(T)$ is zero in both weak and strong coupling limits. This can be verified by substituting the state $\hat \zeta_S(\beta)$ and the term $\hat E^*_S$ in Eq.~\eqref{eq_quantum_uncertainty}, where the commutator between $\hat E^*_S$ and $\hat \zeta_S(\beta)$ becomes zero for this model. The term $\langle\partial_T \hat E^*_S\rangle$ is non-zero for strong coupling but vanishes in the weak coupling limit. The generalized fluctuation-dissipation relation, Eq.~\eqref{specific heat}, is satisfied.
Further, $C_S(T)$ is negative in the strong-coupling regime, which was also observed in~\cite{Campisi2009} and could be ascribed to the factor $\langle\partial_T \hat E^*_S\rangle$ going negative in the strong coupling regime for the same range of temperature. When the specific heat capacity turns negative, it suggests that there exist temperature intervals where the system's energy rises as the temperature dips. This phenomenon might occur because the process of coupling the system to the reservoir may actually decrease the overall heat capacity due to the presence of non-vanishing interactions~\cite{CAMPISI2010187, ingold2009specific}.
Also, $\Delta U_S(T)$ increases steeply with an increase in temperature in both the strong and the weak coupling regimes and approaches a common steady value at higher temperatures.

\begin{figure}
    \centering
    \includegraphics[width=1\linewidth]{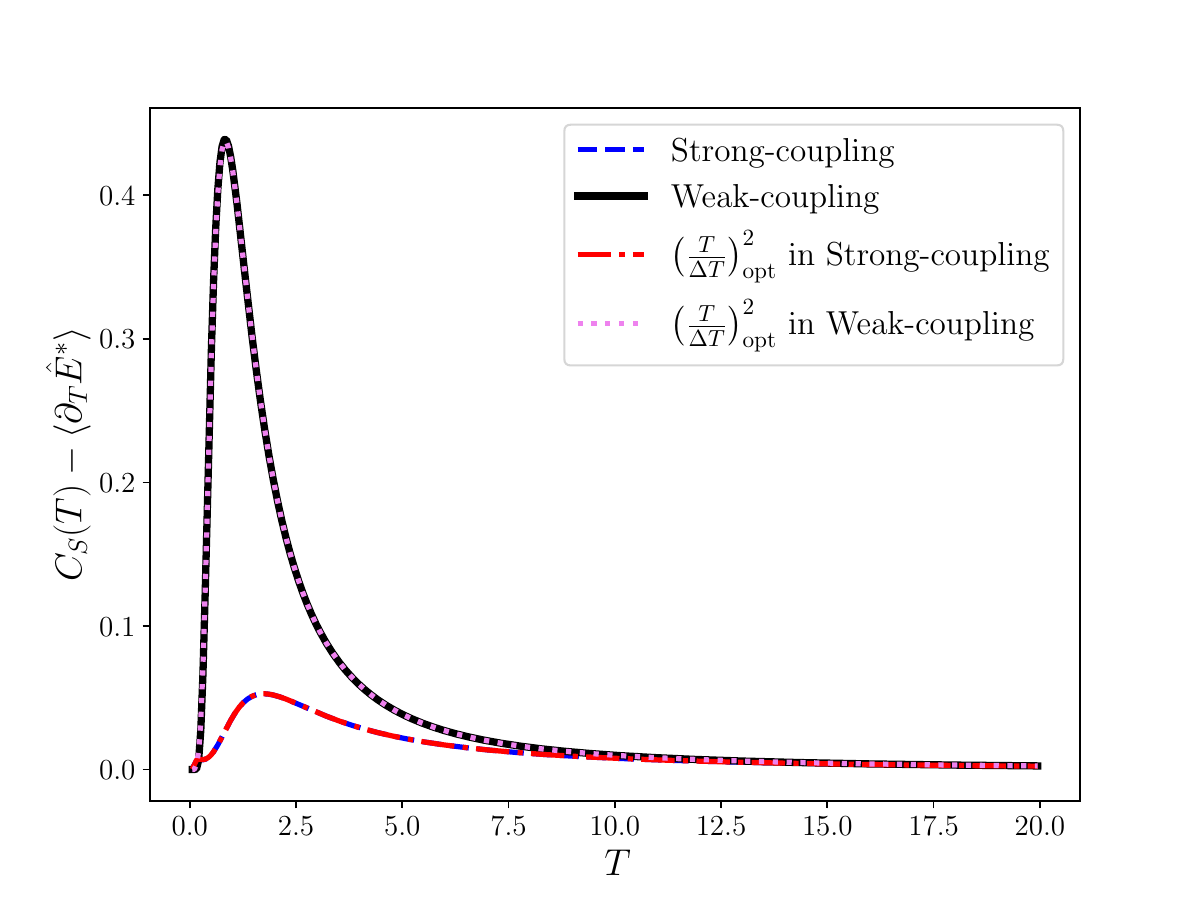}
    \caption{Variation of the the upper bound on the signal-to-noise ratio ($C_S(T) - \langle\partial_T\hat E^*_S\rangle$) for the JC model using Eq.~\eqref{specific heat} as a function of temperature. The parameters are: $\omega_0 = 2.0, \omega_c = 1.0$.}
    \label{fig_upper_bound_nmgad_model_1}
\end{figure}%
The upper bound on the signal-to-noise ratio is depicted in Fig.~\ref{fig_upper_bound_nmgad_model_1} and is observed to be tight both for the weak and strong coupling regimes. This bound is tight due to the diagonal structure of the HMF (consistent with the behavior observed in the two-qubit case). For any value of $p$ and $q$, the Fisher information $F(\beta)$, Eq.~\eqref{qfi}, is given by
\begin{align}
    F(\beta) = \frac{e^{\beta (p + q)}\left[p - q + \beta\left(\partial_\beta p - \partial_\beta q\right)\right]^2}{\left(e^{\beta p} + e^{\beta q}\right)^2},
\end{align}%
which is equal to $\Delta U_S^2$, Eq.~\eqref{eq_delta_US}, for the single qubit model. Further, since the quantum uncertainty is zero, the value of $\left(\frac{T}{\Delta T_S}\right)^2_{\rm opt} = T^2F_S(T)$ becomes equal to $C_S(T) - \langle\partial_T\hat E^*_S\rangle = \frac{\Delta U_S^2}{T^2}$, resulting in the saturation of the bound. 

The ergotropy is calculated using the state $\hat \zeta_S(T) = e^{-\beta \mathcal{H}^*_S}/Z^*_S$, which is zero due to the absence of coherence, and the diagonal terms having ${\rm Tr}[\sigma^z\hat \zeta_S(T)]<0$, that is, the population of the excited state is always lower than the ground state of the system in the state $\zeta_S(T)$~\cite{Tiwari2023}. 

\begin{figure}
    \centering
    \includegraphics[width=1.0\linewidth]{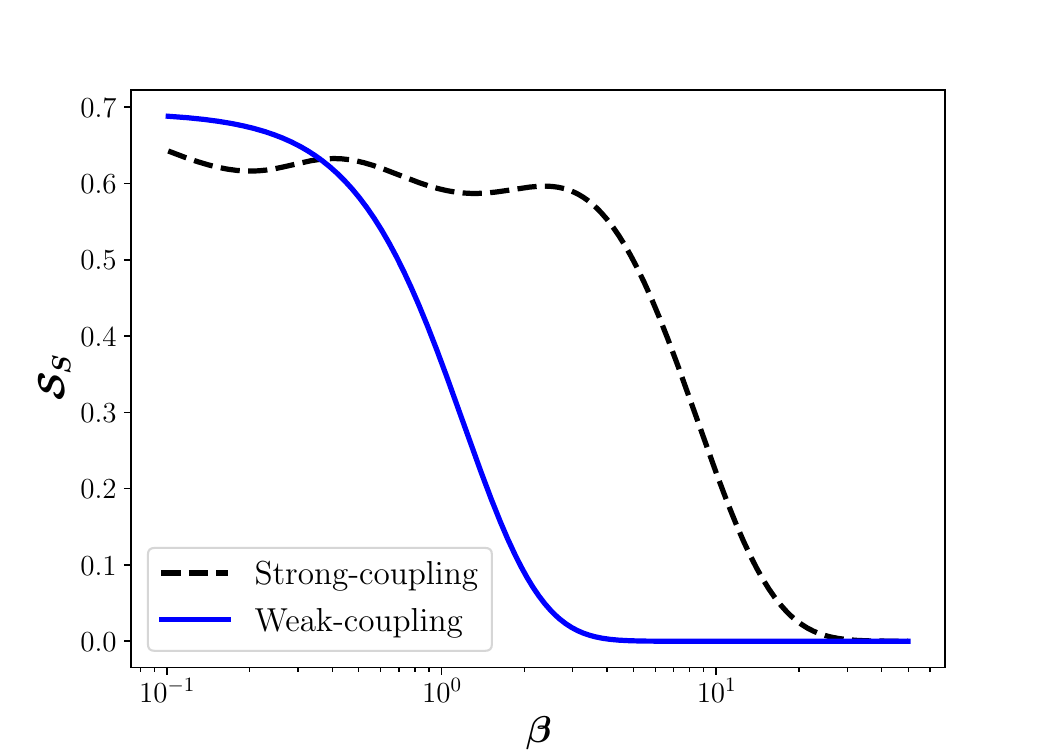}
    \caption{Variation of entropy [Eq.~\eqref{entropy}] as a function of inverse temperature $\beta$ for the JC model. Here, $\omega_0 = 2.0, \omega_c = 1.0$. Strong and weak coupling limits are taken similarly to the previous figures.}
    \label{fig_entropy_nmgad_model}
\end{figure}

To calculate the entropy of the system using the HMF for the JC model, we use Eq.~\eqref{eq_single_qubit_hmf} and substitute it in Eq.~\eqref{entropy}. In Fig.~\ref{fig_entropy_nmgad_model}, we plot the variation of entropy as a function of the inverse temperature $\beta$ of the bath. 
It can be observed that for lower temperatures (higher $\beta$), the entropy $\mathcal{S}_S$ of the system tends to zero, consistent with the third law of thermodynamics. In the strong coupling limit ($\lambda \sim \omega_0$), the entropy of the system becomes non-zero at lower temperatures when compared to the entropy of the system in the weak coupling limit ($\lambda \ll \omega_0$). At higher temperatures, the entropy for weak system-bath coupling takes higher values as compared to the entropy in the strong coupling regime. Also, the entropy shows an oscillatory behavior at higher temperatures in the strong coupling regime. 

\begin{figure}
    \centering
    \includegraphics[width=1\linewidth]{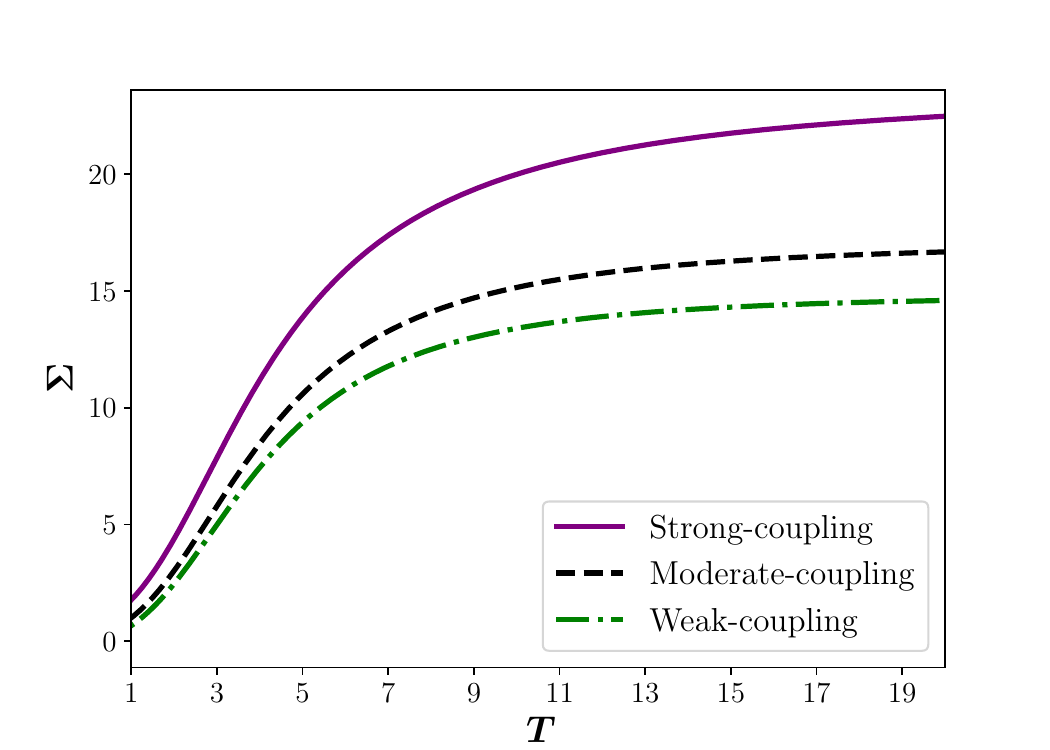}
    \caption{Variation of the entropy production $\Sigma$ as a function of temperature at time $t=0.5$ for the JC model. The parameters are $\omega_0 = 2.0, \omega_c = 1.0$.}
    \label{fig_Entropy_production_nmgad_model}
\end{figure}%
Further, the entropy production for the JC model is calculated using the initial states $\ket{\psi}_S = \frac{1}{\sqrt{2}}\left(\ket{0} + \ket{1}\right)$ and $\rho_B(0) = e^{-\beta H_B}/{\rm Tr}\left[e^{-\beta H_B}\right]$ of the system and the bath, using Eq.~\eqref{eq_entropy_production}, where the unitary evolution is given by $U = e^{-iHt}$. The variation of entropy production as a function of temperature is shown in Fig.~\ref{fig_Entropy_production_nmgad_model}. It exhibits a similar pattern to that of the earlier model, with the strong coupling case attaining higher values. Again, the entropy production shows a saturating behavior with an increase in temperature and remains positive for all values of $T$, consistent with the second law of quantum thermodynamics. The strong-coupling regime and memory effects are advantageous, as can be seen from Figs. \ref{fig_1_qubit_decoherence_model_heat_capacity} and \ref{fig_Entropy_production_nmgad_model}. Together, these observations underscore how strong coupling leads to a resurgence of quantum thermodynamic behavior, typically suppressed in the weak-coupling limit.

\section{Conclusions}\label{conclusions}
In this work, we studied the quantum thermodynamic behavior of open quantum system models. To understand the effect of weak and strong coupling, use was made of the Hamiltonian of Mean Force, which could be interpreted as an effective Hamiltonian of the system instead of the system's bare Hamiltonian in the strong coupling regime. The thermodynamic potentials, particularly specific heat capacity, internal energy, and entropy, were studied using this Hamiltonian of Mean Force.  Further, we also discussed the energy-temperature uncertainty relation that provides an upper bound to the signal-to-noise ratio. This upper bound was computed using the specific heat capacity of the system of interest and a quantity used to compute the internal energy of the system. The behavior of the above thermodynamic quantities was illustrated in two models. In both models, the HMF was seen to have a larger deviation from the system's bare Hamiltonian for strong coupling and low temperature. Further, in the weak coupling regime of both models, quantities such as specific heat capacity, fluctuations in internal energy, quantum uncertainty, and $\left\langle\partial_T\hat E^*_S\right\rangle$ showed similar behavior. However, in the strong coupling regime, the above quantities behaved differently for both models. Particularly, the HMF for the single qubit model did not have off-diagonal terms, and the specific heat capacity in the strong coupling regime had negative values for this model, which was not the case with the two-qubit model. Also, for the single-qubit model, the quantum uncertainty was zero for all the parameter regimes.
First, we considered a two-qubit model interacting with an electromagnetic field, and next, the Jaynes-Cumming model without the rotating wave approximation. The generalized fluctuation-dissipation relation was satisfied for both models for all coupling limits, ranging from weak to strong coupling. The upper bound on the signal-to-noise ratio was tight for the two-qubit model in the strong coupling regime. In the weak coupling regime of the two-qubit model and for the single-qubit model for all coupling strengths, the inequality was saturated, which can be attributed to the diagonal structure of the HMF. Further, in the case of the two-qubit model, the effective Gibbs state based on the HMF had a non-zero capacity for doing work, as its coherent ergotropy was non-zero in strong and moderate coupling limits due to the presence of coherence in this state. The entropy production for both systems was greater in the strong coupling regime, reducing with the decrease in the coupling strength. Our study brings out the impact of non-Markovianity on quantum thermodynamics. For instance, the presence of temperature-dependent corrections to the specific heat, the restoration of coherent ergotropy, and deviations in entropy production, all in the strong coupling regime, demonstrate the resurgence of quantum features typically suppressed in the weak coupling regime. Thus, thermodynamic quantities can serve as indicators of non-Markovian dynamics in quantum systems.
These results would be useful for quantum thermal devices, such as quantum batteries, and provide insights into the complex behavior of open quantum systems in the context of quantum thermodynamics. 

\bibliographystyle{apsrev}
\bibliography{bibliography}

\end{document}